\documentclass[]{revtex4-1}

\usepackage{graphicx}
\usepackage{dcolumn}
\usepackage{bm}
\usepackage[rgb]{xcolor}
\usepackage[normalem]{ulem}
\usepackage{float}
\usepackage{tabularx}
\usepackage{caption}
\usepackage{hyperref}
\usepackage{tikz}
\usepackage{algorithm}
\usepackage{tikz}

\usepackage{amssymb}
\usepackage{amsmath}

\usetikzlibrary{arrows.meta, positioning}

\usepackage{appendix}

\usepackage{color}

\begin{document}


\title{Nonlocal correlations for bosonic fields in black hole quantum atmosphere}


\author{Adam Z. Kaczmarek$^1$}
\author{Johann Gil$^1$}
\author{Zygmunt B{\c a}k$^1$}
\author{Ewa A. Drzazga-Szcz{\c{e}}{\'s}niak$^2$}
\author{Dominik Szcz{\c{e}}{\'s}niak$^1$}
\email{d.szczesniak@ujd.edu.pl}


\affiliation{
${^1}$Institute of Physics, Faculty of Science and Technology, Jan D{\l}ugosz University in Cz{\c{e}}stochowa, 13/15 Armii Krajowej Ave., 42200 Cz{\c{e}}stochowa, Poland\\
${^2}$Department of Physics, Faculty of Production Engineering and Materials Technology, Cz{\c{e}}stochowa University of Technology, 19 Armii Krajowej Ave., 42200 Cz{\c{e}}stochowa, Poland
}

\date{\today}

\begin{abstract}

Recent theoretical studies propose that Hawking radiation may not emerge strictly at the event horizon but rather from the spatially extended region surrounding a black hole, commonly referred to as the quantum atmosphere. In this work, we explore how this concept influences nonlocal quantum correlations in a bosonic bipartite system located at certain distance from a Schwarzschild black hole. By employing the measurement-induced nonlocality (MIN), as a quantifier of quantum correlations, we analyze the response of bosonic fields to the thermal and geometric characteristics associated with the Hartle–Hawking vacuum. In this manner, we extend previous studies that primarily focused on the fermionic systems. Our results reveal that, when quantum atmosphere is taken into account, the behavior of MIN departs from its conventional near-horizon profile. In particular, bosonic nonlocal correlations are found to exhibit a pronounced degradation at a finite radial distance from the event horizon and to ultimately vanish as scaled distance increases further. To some extent this behavior contrasts with the previously considered fermionic case, indicating that bosonic fields provide potentially stronger response to the quantum atmosphere.

\end{abstract}

\maketitle

\section{Introduction}

The interplay between quantum field theory and general relativity has led to some of the profound insights in the modern theoretical physics \cite{birrell1982}. One of the most important result, in this respect, is the prediction of Hawking radiation, according to which black holes (BHs) emit thermal radiation caused by the quantum effects \cite{hawking1974, hawking1975}. In a similar fashion, closely related phenomena also arise in non-inertial frames, not necessarily related to BHs \cite{unruh1976, birrell1982}. Altogether these kind of effects, reveal deep connections between gravitation, thermodynamics, and quantum theory, stimulating an extensive research on the role of quantum information in various gravitational settings \cite{bekenstein1973, hawking1975, wald2001}. This notably contributes to the development of the relativistic quantum information (RQI) framework, which concentrates on the quantum information under the influence of relativistic effects \cite{martinmartinez2011,lanzagorta2014}.

In the realm of BH physics, one of the most persistent and fundamental conceptual challenges challenges concerns both the Hawking radiation and fate of information due to the process of black hole evaporation. This problem is known as the black hole information paradox \cite{hawking1976} and relates to the fact that emitted radiation is purely thermal, appearing to depend only on the macroscopic parameters of a BH, such as its mass, charge, and angular momentum \cite{hawking1975, bekenstein1973, wald2001}. As a result, information about initial quantum state seems to be lost during the evaporation process, violating the principle of unitarity \cite{hawking1976}. Naturally, various solutions to this problem have been proposed, ranging from modifications of semiclassical gravity to more radical scenarios involving firewalls or new quantum gravitational effects \cite{almheiri2013, page2005}. However, before asking whether information is lost, preserved, or recovered it is instructive to answer question concerning the physical origin of Hawking radiation itself. Here, the traditional viewpoint assumes that the radiation is generated extremely close to the event horizon \cite{hawking1975}. Nonetheless, more recent studies suggest that the radiation may actually arise from an extended region surrounding the black hole, often referred to as the quantum atmosphere \cite{giddings2016,hod2016, dey2017}. In this scenario, the effective emission region lies at a finite radial distance ($r$) from the horizon \cite{giddings2016, hod2016} and such argumentation appears to be supported by studies on the stress-energy tensor, Schwinger effect or mass-superposed BTZ BHs \cite{dey2017,dey2019,zhang2025quantum}. Hence, if viable, this concept may significantly affect our understanding of the behavior of quantum systems in relativistic setting, bringing us closer to solution of the black hole information paradox \cite{hod2016, dey2017}.

In respect to the above, a convenient insight into the quantum atmosphere concept can be provided within RQI framework. From this perspective, the quantum correlation measures appear as a particularly powerful tool for probing fundamental aspects of quantum fields in relativistic environments \cite{peres2002}. In particular, measures such as entanglement, quantum discord, and measurement-induced nonlocality (MIN) provide means for investigating structure of quantum states influenced by gravitational backgrounds \cite{peres1996, ollivier2001, luo2011}. In the context of the quantum atmosphere, the previous works were mainly devoted to the study of different correlations and resources, ranging from nonlocality to the coherence \cite{kaczmarek2024,kaczmarek2025,liu2024quantum,zhang2025entanglement,zhang2026quantum,liu2026quantum,ming2026nonlocal}. Among these, MIN has attracted considerable attention as a measure of nonlocal correlations beyond entanglement \cite{luo2011, ming-liang2012}. Introduced from a geometric perspective, MIN is a quantifier of the maximum global effect caused by locally invariant measurements \cite{luo2011}. In contrast to the closely related geometric quantum discord, the MIN is defined over all the von Neumann measures that do not disturb the local state \cite{luo2011, ollivier2001, dakic2010}. However, although certain signatures of quantum atmosphere has been reveled within MIN spectrum, the corresponding discussion was limited only to the fermionic modes due to their robust characteristics under the influence of the external radiation \cite{kaczmarek2023,kaczmarek2024,zhang2025entanglement,zhang2026quantum,liu2026quantum,ming2026nonlocal}. In particular, fermions are protected by the Pauli exclusion principle, whereas vastly unexplored bosonic fields allow for an unbounded occupation numbers. Hence, it is argued here that this bosonic character may lead to a stronger degradation mechanisms and somewhat different behavior of quantum correlations. This motivates investigations of bosonic correlation as well, not only as a straightforward extension but an important step toward better understanding of quantum atmosphere signatures in correlations spectrum.

In particular, here we investigate nonlocal quantum correlations for bosonic fields in the vicinity of a Schwarzschild black hole, by employing MIN measure as a quantifier of quantum correlations in a bipartite system. In this framework, we analyze how the thermal properties associated with the Hartle-Hawking vacuum influence the behavior of bosonic quantum correlations \cite{hartlehawking1976, tolman1930}. The considered bipartite setup involves two observers, Alice and Bob, who share a maximally entangled Bell state in the asymptotically flat region of spacetime. While Alice remains far from the black hole, Bob moves toward it and probes the field in its curved spacetime. This allows us to explore how the scaled radial distance $r/r_{H}$ (where $r_{H}$ denotes the horizon radius), the scaled Hartle-Hawking temperature $T_{HH}/T_{H}$ (where $T_{HH}$ is the bare Hartle-Hawking temperature and $T_{H}$ denotes the Hawking temperature \cite{kaczmarek2023}), and the Hartle-Hawking constant $D_{HH}$ (controls the strength of the thermal effects in the quantum atmosphere picture \cite{eune2019}) shape the behavior of bosonic quantum correlations. Our goal is to determine whether the presence of a quantum atmosphere modifies the spatial profile of these correlations. 

The work is organized as follows: in Section \ref{2}, we derive the vacuum structure of the bosonic fields in a black hole spacetime. Moreover, we derive the MIN for the bosonic case and introduce the expression of $T_{HH}(r)$, which helps us to calculate the expression of $D_{HH}$. Next, in Section \ref{3}, we discuss how $r/r_{H}$, $T_{HH}/T_{H}$, and $D_{HH}$ influence the behavior of the bosonic MIN. Finally, we conclude our discussion with some pertinent remarks and perspectives for future research.

\section{Theoretical background}
\label{2}

In the present section we provide the theoretical background required to describe bosonic fields in a black hole spacetime and the associated nonlocal correlations. First, we consider a massless bosonic scalar field in Schwarzschild spacetime, described by the Klein-Gordon equation. This allows us to derive vacuum structure in the presence of an event horizon. Next, we introduce MIN for the bosonic case, which will be used as the main quantifier of quantum correlations throughout this work. The local Hartle-Hawking temperature is also given explicitly along with the expression of the Hartle-Hawking parameter. This framework allow us to discuss in Section \ref{3} how the bosonic MIN varies with $r/r_{H}$, $T_{HH}/T_{H}$, and $D_{HH}$, providing insights into the nontrivial effects of a spatially dependent thermal environment on bosonic quantum correlations. 

\subsection{Bosonic fields in a black hole spacetime}

To study the behavior of bipartite coherence near the quantum atmosphere region, we begin with the initial bosonic state. In the first step, massless bosonic scalar field $\phi(x)$ satisfying the Klein-Gordon equation for the curved spacetime is considered \cite{navarro2005, pan2008}:
\begin{equation}
\Box \phi = \frac{1}{\sqrt{-g}} \partial_\mu (\sqrt{-g} g^{\mu\nu} \partial_\nu \phi) = 0,
\label{KGeq}
\end{equation}
where $g_{\mu\nu}$ is the Schwarzschild metric and $g = \det(g_{\mu\nu})$. 

Following Gidding's argument \cite{giddings2016}, we consider the dimensionally reduced line element for a Schwarzschild black hole:
\begin{equation}
ds^2 = -f(r) dt^2 + f(r)^{-1} dr^2,
\end{equation}
and substitute it into the Klein-Gordon equation \eqref{KGeq}.

In detail, by considering equation \eqref{KGeq} near the event horizon, we can arrive at the incoming wave function, which is analytic for the entire spacetime and given as:
\begin{equation}
\phi^{in} = Y_{lm}e^{i\Omega u}.
\label{(3)}
\end{equation}
Similarly, the outgoing wave functions for the regions I and II of the event horizon are \cite{pan2008}:
\begin{equation}
\phi_{\Omega}^{I} = Y_{lm}e^{i\Omega u}, \ \phi_{\Omega}^{II} = Y_{lm}e^{-i\Omega u}.
\label{(4)}
\end{equation}
In equations \eqref{(3)} and \eqref{(4)}, $\Omega$ is the bosonic frequency of the mode with the units $1/l_{p}$ where $l_{p}$ is the Plank length \cite{kaczmarek2023} and $Y_{lm}(\theta,\varphi)$ denotes scalar spherical harmonics. Analogously to the fermionic case, the $\phi^{I}$ and $\phi^{II}$ solutions span orthogonal basis for the Klein-Gordon field \cite{pan2008}:
\begin{equation}
\phi^{out} = \sum_{lm} \int d\Omega \, [ a_{\Omega}^{I} \phi_{\Omega}^{I} + a_{\Omega}^{II} \phi_{\Omega}^{II} + b_{\Omega}^{\dagger I} \phi_{\Omega}^{* I} + b_{\Omega}^{\dagger II} \phi_{\Omega}^{* II} ], 
\label{(5)}
\end{equation}
where $a_{\Omega}^{I}$ ($a_{\Omega}^{II}$) and $b_{\Omega}^{\dagger I}$ ($b_{\Omega}^{\dagger II}$) are the bosonic annihilation and creation operators for the exterior (interior) region of the considered BH. Note again that the summation in equation \eqref{(5)} is over the spherical harmonics (see equation \eqref{(4)}). 

By introducing the following abbreviations:
\begin{equation}
cosh \ r = \frac{1}{\sqrt{1-e^{-\Omega/T}}}, \ tanh \ r = e^{-\Omega/T},
\end{equation}
and
\begin{equation}
tanh \ r = t(T), \ cosh \ r = 1/(\sqrt{1-t(T)^{2}}),
\end{equation}
we can write down the vacuum and excited bosonic modes in a more convenient form:
\begin{equation}
|0\rangle_{k} = \sqrt{1-t(T)^{2}} \sum_{n=0}^{\infty} t(T)^{n} |n\rangle_{I} |n\rangle_{II}, \ |1\rangle_{k} = (1-t(T)^{2}) \sum_{n=0}^{\infty} \sqrt{n+1} t(T)^{n} |n+1\rangle_{I} |n\rangle_{II},
\label{STATES}
\end{equation}
where $T$ is the  temperature. In equation \eqref{STATES}, the presence of the infinite sum reflects the unbounded occupation number of bosonic modes. The equation \eqref{STATES} correspond to decomposition of the bosonic modes in the Kruskal basis.

The nontrivial structure of the vacuum, together with the presence of the horizon, leads to correlations between modes in regions I and II. These correlations, encoded in the two-mode structure of the bosonic states, will be analyzed in terms of MIN in the next subsection.

\subsection{Nonlocal correlations in the Hartle-Hawking vacuum}

Here, we employ a conventional setup that follows scenarios originally presented in \cite{fuentes2005, he2016, ge2008}. To be more precise, two observers are considered.  First is Alice equipped with the particle detector sensitive only to mode $|n\rangle_{A}$ and second is Bob detecting only mode $|n\rangle_{B}$ by using his device. In such a framework, both observers initially share a maximally entangled Bell state (this is equivalent to setting the entanglement parameter $\eta = 1$) for the same event in the flat region of Minkowski spacetime. Next, it is assumed that Alice remains stationary at the asymptotically flat region of the spacetime $(r \rightarrow \infty)$, while Bob first freely falls in the direction of a black hole and at some point starts to hover near the event horizon. As a result, Bob becomes affected by the thermal bath of particles associated with the Hartle-Hawking vacuum. In the quantum atmosphere picture, the effective temperature of this bath depends on the radial distance from the black hole, and thus it is not constant. Therefore, to describe what Bob will detect, the mode $|n\rangle_{B}$ is given in the coordinates of a black hole. 

Here, it is instructive to explain several aspects of the above setup. First, since entanglement is sensitive to the environment, we note that Bob's trajectory may alter the description of the state. To avoid this potential issue, it is assumed that Bob is free-falling toward a black hole, slowly decelerating before becoming stationary \cite{ge2008}. Second, we adopt the single-mode approximation for the construction of the bosonic bipartite states \cite{alsing2003}. Despite its limitations, such an approach has been found to capture the most important qualitative features of the correlations of more sophisticated models \cite{bruschi2010}. To this end, one can find additional details of the method employed here in previous related studies \cite{he2016, pan2008, kaczmarek2023, HU2018}.  

The analysis of the MIN for bosons starts with the following Bell state \cite{zettili2009quantum}:
\begin{equation}
    \rho_{AB}=|\phi^{+}\rangle \langle \phi^{+}|,
\end{equation}
for
\begin{equation}
|\phi^{+}\rangle = \frac{1}{\sqrt{2}} (|00\rangle + |11\rangle). 
\label{Bell}
\end{equation}
By using equation \eqref{STATES}, one can modify equation \eqref{Bell} as:
\begin{equation}
|\phi^{+}\rangle = \frac{1}{\sqrt{2}} (\sqrt{1-t(T)^{2}} \sum_{n=0}^{\infty} t(T)^{n} |0,n_{I},n_{II}\rangle  + (1-t(T)^{2}) \sum_{n=0}^{\infty} \sqrt{n+1} t(T)^{n} |1,n+1_{I},n_{II}\rangle). 
\end{equation}
Note, that bosonic fields allow for arbitrary occupation numbers, the expansion involves an infinite sum over number of states. As a consequence, the resulting bipartite state {\it lives} in an infinite-dimensional Hilbert space, which significantly increases the complexity of the associated density matrices. In contrast, for fermionic fields the Pauli exclusion principle restricts the occupation number, leading to a finite-dimensional Hilbert space. Thus, extra care is taken when dealing with bosonic modes within our framework \cite{richter2015degradation,kaczmarek2023}.

Since the modes in region II are causally disconnected from an observer confined to region I due to the presence of the horizon, they are physically inaccessible. Therefore, the appropriate physical description of the system is obtained by tracing out these degrees of freedom. The resulting reduced density matrix is given by:
\begin{equation}
\rho_{{AB}_{I}} = Tr_{B_{II}}(\rho_{AB_{I}B_{II}}).
\end{equation}
This procedure effectively removes the inaccessible degrees of freedom and transforms the global pure state into a mixed one, reflecting the loss of information associated with region II. As a result, the initially tripartite system composed of $A$, $B_{I}$, and $B_{II}$ is reduced into an effective bipartite system involving only $A$ and $B_{I}$, on which the MIN can be consistently evaluated \cite{fuentes2005, he2016}.

Within this framework, the measurement-induced nonlocality (MIN) of the bipartite quantum state $\rho$ shared by $A$ and $B$ subsystems is defined as \cite{luo2011}:
\begin{align}
   \text{MIN}(\rho)=\text{Max}_{\Pi^ A}\mid\mid \rho- \Pi^A (\rho)\mid\mid^2.
   \label{mindef}
\end{align}
In this context, to study the MIN for the bosonic state given by the $\rho_{AB_{I}}$, we need to use projective measurements on Alice's state. Such measurements can be taken over the single-qubit system. Thus, it is possible to parametrize it by the unit vector $\vec{x} = (x_{1},x_{2},x_{3})$ with the aid of the projectors \cite{brown2012}:
\begin{equation}
\Pi_{\pm} = \frac{1}{2} (I \pm \vec{x} \cdot \vec{\sigma}).
\end{equation}
Next, these projectors can be rewritten in the following form:
\begin{equation}
\Pi_{\pm} = \frac{1}{2} [(1 \pm x_{3}) |0\rangle \langle0| + (1 \mp x_{3}) |1\rangle \langle1| \pm (x_{1} - ix_{2}) |0\rangle \langle1| \pm (x_{1} + ix_{2}) |1\rangle \langle0|].
\label{(12)}
\end{equation}
Based on equation \eqref{(12)}, one can obtain the post-measured final state:
\begin{equation}
\rho_{AB_{I}}^{'} = \sum_{\alpha = \pm} (\Pi_{\alpha} \otimes I) \rho_{AB_{I}} (\Pi_{\alpha} \otimes I) = \sum_{\alpha = \pm} p_{\alpha} \Pi_{\alpha} \otimes \rho_{B_{I}|\alpha},
\label{DISCORD}
\end{equation}
where:
\begin{equation}
\rho_{B_{I}|\alpha} \equiv Tr ((\Pi_{\alpha} \otimes I_{B}) \rho_{AB_{I}} (\Pi_{\alpha} \otimes I_{B}))/p_{\alpha},
\end{equation}
is the post-measured state of Bob's system conditioned on the outcome $\alpha$ with the probability $p_{\alpha}$. Moreover, with the aid of equation \eqref{(12)}, we can obtain ($p_{\pm} = 1/2$) \cite{brown2012}:
\begin{equation}
\rho_{B_{I}|\pm} \equiv \frac{1-t(T)^{2}}{2} \tilde{\rho}_{I},
\end{equation}
where:
\begin{equation}
\tilde{\rho}_{I \pm} = (1 \pm x_{3})M_{00} + (1 \mp x_{3})M_{11} \pm (x_{1} - ix_{2})M_{01} \pm (x_{1} + ix_{2})M_{10},  
\end{equation}
and with the following matrices for Bob's Hilbert space:
\begin{equation}
\begin{aligned}
M_{00} &= \frac{1}{2} (\eta + 1) t(T)^{2n} |n\rangle \langle n|, \\
M_{10} &= \eta \sqrt{n+1} \sqrt{1 - t(T)^{2}} t(T)^{2n} |n+1\rangle \langle n|, \\
M_{01} &= M_{10}^{\dagger}, \\
M_{11} &= \frac{1}{2} (\eta + 1) (n + 1) (1 - t(T)^{2}) t(T)^{2n} |n+1\rangle \langle n+1|.
\end{aligned}
\end{equation}
After the above preparations, the quantity of interest can now be calculated:
\begin{equation}
Tr((\rho_{AB_{I}} - \rho_{AB_{I}}^{'})^{2}) = \frac{(1 - t(T)^{2})^{2}}{4} [Tr(X_{00}^{2}) + 2 \ Tr(X_{01}X_{10}) + Tr(X_{11}^{2})],
\end{equation}
with the X's given by:
\begin{equation}
\begin{aligned}
X_{00} &\equiv M_{00} - \frac{1}{4} [(1 + x_{3}) \tilde{\rho}_{+} + (1 - x_{3}) \tilde{\rho}_{-}], \\
X_{11} &\equiv M_{11} - \frac{1}{4} [(1 - x_{3}) \tilde{\rho}_{+} + (1 + x_{3}) \tilde{\rho}_{-}], \\
X_{01} &\equiv M_{01} - \frac{1}{4} (x_{1} - ix_{2}) (\tilde{\rho}_{+} - \tilde{\rho}_{-}), \\
X_{10} &\equiv M_{10} - \frac{1}{4} (x_{1} + ix_{2}) (\tilde{\rho}_{+} - \tilde{\rho}_{-}).
\end{aligned}
\end{equation}
We note that traces of the X's will be given by the linear combinations of the matrices M, reducing the quantity of interest $Tr((\rho_{AB_{I}} - \rho_{AB_{I}}^{'})^{2})$ to \cite{brown2012}:
\begin{equation}
Tr((\rho_{AB_{I}} - \rho_{AB_{I}}^{'})^{2}) = \frac{(1 - t(T)^{2})^{2}}{8} [(1 - x_{3}^{2})(Tr(M_{00}^{2}) + Tr(M_{11}^{2}) - 2 \ Tr(M_{00}M_{11}) + 2(1 + x_{3}^{2}) \ Tr(M_{01}M_{10}))],
\end{equation}
with their traces equal to:
\begin{equation}
\begin{aligned}
Tr(M_{00}^{2}) &= \frac{1}{4} (1+\eta)^{2} \sum_{n=0}^{\infty} t(T)^{4n}, \\
Tr(M_{01}M_{10}) &= \frac{\eta^{2}(1-t(T)^{2})}{(t(T)^{4}-1)^{2}}, \\
Tr(M_{11}^{2}) &= - \frac{(\eta+1)^{2}(t(T)^{4}+1)}{4(t(T)^{2}-1)(t(T)^{2}+1)^{3}}.
\end{aligned}
\end{equation}
For the $x_{3} = 1$, the MIN for the bosonic case can finally be obtained:
\begin{equation}
MIN(\rho_{AB_I})=Tr((\rho_{AB_{I}} - \rho_{AB_{I}}^{'})^{2}) = \frac{\eta^{2}(1-t(T)^{2})^{2}(t(T)^{2}-1)^{3}(t(T)^{4}+1)}{8(t(T)^{4}-1)^{3}}.
\end{equation}
We remark that setting $x_{3} = 0$ in equation \eqref{DISCORD} leads to the quantum discord \cite{tian2013, brown2012}.

We note that, since our goal is to characterize possible signatures of the quantum atmosphere on a bipartite coherence, the local temperature (T) should take the Hartle-Hawking ($T_{HH}(r)$) form \cite{eune2019, kaczmarek2024}:
\begin{equation}
T_{HH}(r) = T_{H} \sqrt{1 - \frac{r_{H}}{r}} \sqrt{1 + 2 \frac{r_{H}}{r} + (\frac{r_{H}}{r})^2 (9 + 4D_{HH} + 36 ln(\frac{r_{H}}{r}))},
\label{THH}
\end{equation}
with $T_{H} = \frac{1}{4 \pi r_{H}}$. In equation \eqref{THH}, the $D_{HH}$ is the undetermined constant of the stress tensor for the Hartle-Hawking vacuum, called here the Hartle-Hawking constant for simplicity. According to \cite{eune2019}, its value cannot be fixed by the Hartle-Hawking boundary conditions. However, it is known that for $D_{HH} \ge D_{C} \approx 23.03$ the temperature is positive and decreases after reaching peak at $r_{c} \approx 1.43 r_{H}$. Note that this local temperature is vanishing at the horizon $r_{H}$ and approaches Hawking temperature at the infinity $(r \rightarrow \infty)$. As noticed by Eune and Kim \cite{eune2019}, the temperature \eqref{THH} near horizon limit differs from the conventional Tolman one ($T = T_{H} / \sqrt{f(r)}$) \cite{tolman1930}. Moreover, this relation is a temperature of the outgoing Hawking radiation for a considered scenario \cite{eune2019}. As such, the $T_{HH} = 0$ result agrees with the lack of the influx/flux of the particle radiation for the horizon in thermal equilibrium \cite{eune2019}. At the same time, $T_{HH}$ peaks at the macroscopic distance from the event horizon where the main excitations occur. Note that temperature \eqref{THH} aligns with the main motives of Giddings work \cite{giddings2016, eune2019}. 

It is worth to note that one can in principle attempt to obtain the real value of constant $D_{HH}$, since knowing the ratios $r/r_H$ and $T_{HH}/T_H$ can help with the calculation of $D_{HH}$ via equation (\ref{THH}):
\begin{align}
    D_{HH}= \frac{1}{4} \left(-r/r_H ^2-2 r/r_H -36 \log \left(\frac{1}{r/r_H }\right)+\frac{r/r_H ^3 T_{HH}^2}{(r/r_H -1) T_H^2}-9\right).
\end{align}
In summary, equation \eqref{THH} is pivotal to our analysis since it allows us to examine sensitivity of quantum correlations in a bipartite system to the local temperature variations with distance. As a result, it is possible to search for the signatures of quantum atmosphere in these correlations.     

\section{Thermal evolution}
\label{3}

\begin{figure}
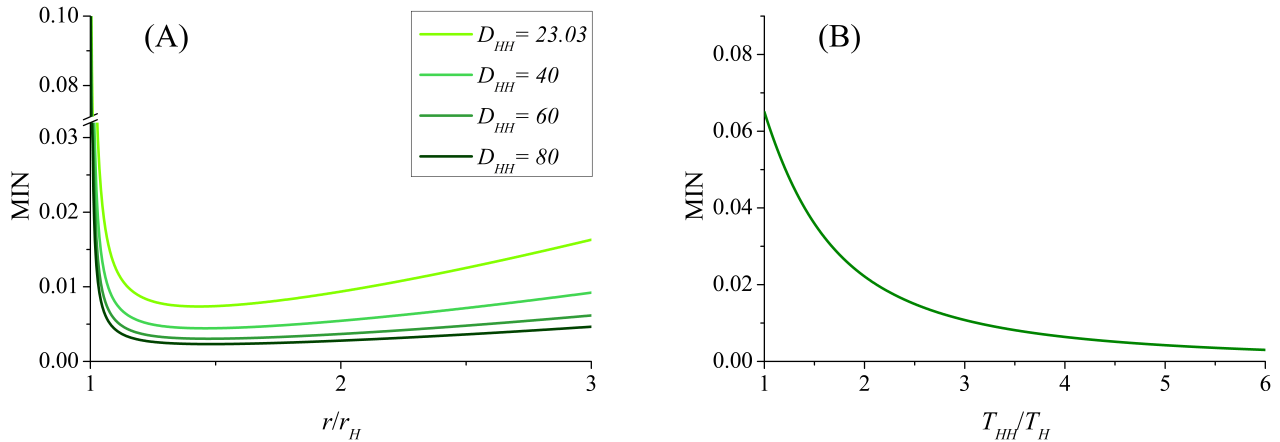

\begin{minipage}{.5\textwidth}
        \centering
        \includegraphics[width=0.98\linewidth]{fig01a.png}
    \end{minipage}%
    \begin{minipage}{0.5\textwidth}
        \centering
        \includegraphics[width=0.98\linewidth]{fig01b.png}
    \end{minipage}
   \caption{(A) The measurement-induced nonlocality (MIN) for the physically accessible bosonic quantum modes as a function of the scaled distance $(r/r_H)$ and (B) as a function of scaled temperature $T_{HH}/T_H$, for the selected values of the so-called Hartle-Hawking constant ($D_{HH}$) and $\Omega=1$.}
    \label{fig:1}
\end{figure}

We begin our analysis by examining the behavior of nonlocal correlations, via MIN measure, for physically accessible correlations in the bosonic $\rho_{AB}$ state. In Fig. (\ref{fig:1}) (A) the $MIN(\rho_{AB_{I}})$ is presented as a function of the normalized radius $r/r_H$ for selected values of $D_{HH}$. Note, that we are considering $D_{HH}\geq D_C \approx 23.03$, in order to obtain only physically relevant solutions \cite{eune2019}. Since there is no particle flux at the location of horizon, the MIN reaches maximum value when $r/r_H \rightarrow 1$. Afterwards, MIN decreases rapidly, reaching minimum value for radii corresponding to the region well outside the horizon, in the range $r/r_H \in [1.43,1.51]$. Moreover, as radius becomes larger, i.e. system is used as probe at different places, the MIN is still vanishingly small, despite slow value recovery. This is a behavior expected for MIN of bosons within standard BH picture \cite{kaczmarek2023}. In this sense, once the particle flux at some distance $r$ away from horizon influences the system, the nonlocal correlations are  destroyed and can never fully recover to their state before entering the region where energy fluxes originate. It means that extra care should be taken when considering bosons in shaping nonlocality as a resource within the quantum atmosphere framework developed by Eune and Kim \cite{eune2019, kaczmarek2024}. This is confirmed when inspecting Fig. (\ref{fig:1}) (B), where MIN is presented as a function of scaled temperature $T_{HH}/T_H$. The MIN decreases whenever temperature rises, as $\lim_{T_{HH}/T_H}MIN\rightarrow0$. From this perspective, in order to verify whether or not quantum atmosphere region influences considered correlations, the exact location of the subsystem $A$ is needed. This is required since treating MIN as a function of temperature alone, may not be sufficient. Just like for the nonlocal correlations for fermions, the minimal values are independent and are not shifting as the $D_{HH}$ changes \cite{kaczmarek2024}. This is in accordance with the scaled temperature behavior, where the minimal value corresponds to $r/r_H \approx 1.43$. 

\begin{figure*}
    \centering
    \includegraphics[scale=0.8]{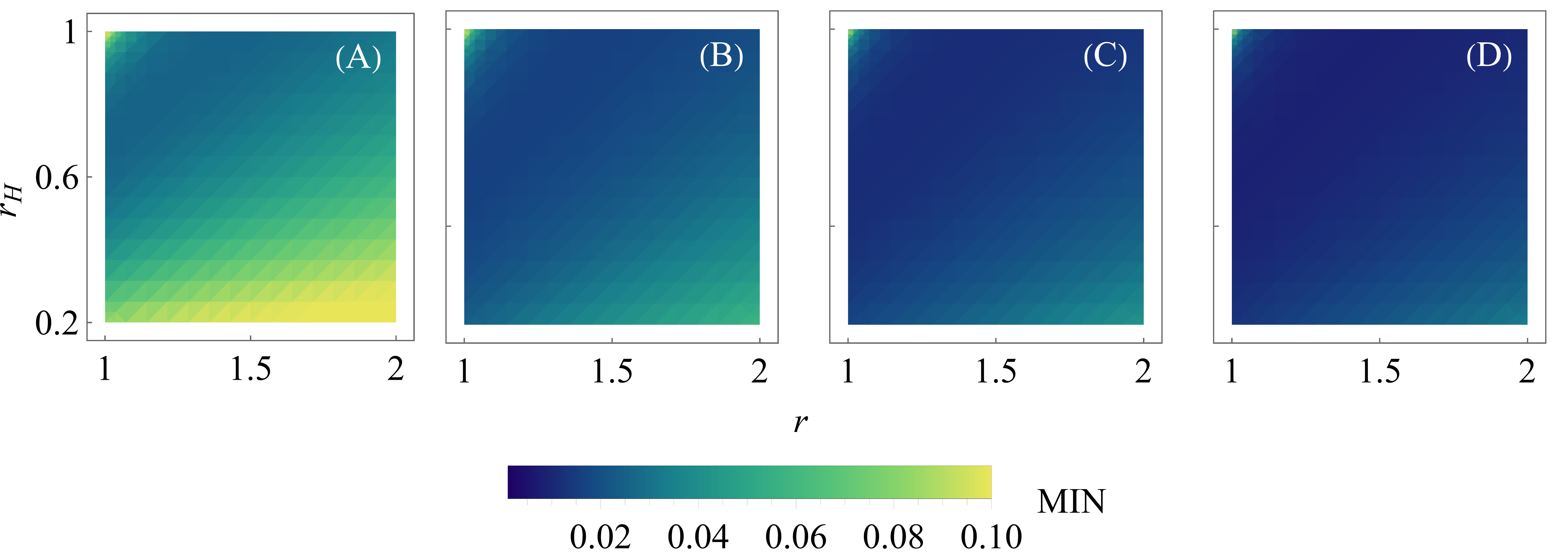}
    \caption{(A)-(D) The measurement-induced nonlocality of the physically accessible bosons ($\text{MIN}(\rho_{AB_I})$) on a plane characterized by the distance from a black hole ($r$) and the event horizon size ($r_{H}$). The results are depicted for the set of selected values of constant ($D_{HH} \in \{23.03,40,60,80\}$). The diagrams are plotted for the $\Omega=1$.}
    \label{fig2}
\end{figure*}

To further verify this observations for different horizon radii, the bosonic MIN measure is next analyzed from the perspective $r$ and $r_H$ separately, as presented in Fig. \ref{fig2}. The diagrams were obtained for the fixed values of constant $D_{HH} \in \{23.03,40,60,80\}$, respectively. Comparing panels (A)-(D) of Fig. \ref{fig2}, we observe that the increase of $D_{HH}$ from the critical value $D_C \approx 23.03$ to $80$, systematically reduces the spatial extent of the region where the MIN measure attains larger values. The shrinkage of this region on the $(r, r_H)$ plane, suggests that parameter $D_{HH}$ controls not only the existence of physical solutions but also the size of the quantum atmosphere where nonlocal correlations remain affected by the temperature. This is particularly visible for panel (D), that corresponds to the $D_{HH}=80$, where the MIN is almost order of magnitude smaller than the one given for the $D_{HH}=23.03$. Another notable observation is the non-monotonic behavior along constant-$r$ slices. Specifically for fixed $r$, MIN initially decreases with increasing $r_H$, reaches a minimum and then remains at near-zero values. This threshold behavior may suggest the existence of a critical horizon size above which the destructive role of quantum atmosphere on bosonic nonlocal correlations becomes even more evident, regardless of the probe's radial position. This complements the previous analysis of fermionic fields, showing that signatures of the existence of the quantum atmosphere can also be observed from the perspective of the quantum correlation measure considered here \cite{kaczmarek2024}.

\section{Summary and conclusions}

In this paper, we have focused on the analysis of the BH radiation influence on the bosonic MIN quantum measure, in the context of the BH quantum atmosphere argument \cite{giddings2016}. After the description of the bosonic bipartite system in the Hartle-Hawking vacuum, we have analyzed distinct characteristics and features of the nonlocal correlations, as quantified by the MIN \cite{luo2011}. Our analysis shown that in the case of bosons, the evolution of the MIN in the physically accessible correlation can indeed be influenced by the change in the radiation origin, quanitfied by the influence of the local temperature $T_{HH}$. Specifically, the continuous evolution is affected by the change in the normalized distance $r/r_H$ from the event horizon, as sudden reduction of the MIN is observed at $r/r_H \in [1.43,1.51]$. After that the correlations never recover again and practically vanishes as scaled distance grows larger. This observation has particularly important implication for detection of the signatures of BH quantum atmosphere. In particular, for a given black hole of radius $r_H$, there exist an optimal radial window, where bosonic probes can extract nonlocal correlations before they are destructed by the produced radiation. Outside this window, i.e. after some critical $r$, the MIN effectively vanishes, rendering bosonic systems useless as a resources for quantum tasks in quantum atmosphere region. 

The above observations coincide with the behavior obtained when standard description of the Hawking radiation origin is considered \cite{kaczmarek2023}. At the same time, it is somewhat different from the behavior observed for the fermionic-based systems \cite{kaczmarek2024,liu2026quantum,ming2026nonlocal}. This difference can be attributed to fundamental statistical differences in thermal excitations. In detail, fermionic modes, constrained by the Pauli exclusion principle, exhibit saturation effect that limits thermal noise, while bosonic ones undergo Bose statistics, amplifying radiation effects and accelerating loss of correlations within the system of interest \cite{richter2015degradation,kaczmarek2023}. Thus, fermionic detectors not only may retain correlations at larger distances but also exhibit more gradual degradation profile, making them more suitable for reconstruction and tests of the quantum atmosphere structure \cite{kaczmarek2024,kaczmarek2025,ming2026nonlocal}. As such, the presented results complements previous analysis of the fermionic fields and show that signatures of quantum atmosphere depend strongly on quantum statistics of the field. In what follows, the bosonic systems may provide more sensitive probe systems than fermionic ones, providing additional non-trivial insight into the quantum atmosphere argument and near-horizont physics of a BHs.

\bibliographystyle{apsrev}
\bibliography{manuscript}

\begin{thebibliography}{45}
\expandafter\ifx\csname natexlab\endcsname\relax\def\natexlab#1{#1}\fi
\expandafter\ifx\csname bibnamefont\endcsname\relax
  \def\bibnamefont#1{#1}\fi
\expandafter\ifx\csname bibfnamefont\endcsname\relax
  \def\bibfnamefont#1{#1}\fi
\expandafter\ifx\csname citenamefont\endcsname\relax
  \def\citenamefont#1{#1}\fi
\expandafter\ifx\csname url\endcsname\relax
  \def\url#1{\texttt{#1}}\fi
\expandafter\ifx\csname urlprefix\endcsname\relax\def\urlprefix{URL }\fi
\providecommand{\bibinfo}[2]{#2}
\providecommand{\eprint}[2][]{\url{#2}}

\bibitem[{\citenamefont{Birrell and Davies}(1982)}]{birrell1982}
\bibinfo{author}{\bibfnamefont{N.~D.} \bibnamefont{Birrell}} \bibnamefont{and}
  \bibinfo{author}{\bibfnamefont{P.~C.~W.} \bibnamefont{Davies}},
  \emph{\bibinfo{title}{Quantum Fields in Curved Space}}
  (\bibinfo{publisher}{Cambridge University Press}, \bibinfo{year}{1982}).

\bibitem[{\citenamefont{Hawking}(1974)}]{hawking1974}
\bibinfo{author}{\bibfnamefont{S.~W.} \bibnamefont{Hawking}},
  \bibinfo{journal}{Nature} \textbf{\bibinfo{volume}{248}}, \bibinfo{pages}{30}
  (\bibinfo{year}{1974}).

\bibitem[{\citenamefont{Hawking}(1975)}]{hawking1975}
\bibinfo{author}{\bibfnamefont{S.~W.} \bibnamefont{Hawking}},
  \bibinfo{journal}{Communications in Mathematical Physics}
  \textbf{\bibinfo{volume}{43}}, \bibinfo{pages}{199} (\bibinfo{year}{1975}).

\bibitem[{\citenamefont{Unruh}(1976)}]{unruh1976}
\bibinfo{author}{\bibfnamefont{W.~G.} \bibnamefont{Unruh}},
  \bibinfo{journal}{Physical Review D} \textbf{\bibinfo{volume}{14}},
  \bibinfo{pages}{870} (\bibinfo{year}{1976}).

\bibitem[{\citenamefont{Bekenstein}(1973)}]{bekenstein1973}
\bibinfo{author}{\bibfnamefont{J.~D.} \bibnamefont{Bekenstein}},
  \bibinfo{journal}{Physical Review D} \textbf{\bibinfo{volume}{7}},
  \bibinfo{pages}{2333} (\bibinfo{year}{1973}).

\bibitem[{\citenamefont{Wald}(2001)}]{wald2001}
\bibinfo{author}{\bibfnamefont{R.~M.} \bibnamefont{Wald}},
  \bibinfo{journal}{Living Reviews in Relativity} \textbf{\bibinfo{volume}{4}}
  (\bibinfo{year}{2001}).

\bibitem[{\citenamefont{Martin-Martinez}(2011)}]{martinmartinez2011}
\bibinfo{author}{\bibfnamefont{E.}~\bibnamefont{Martin-Martinez}},
  \bibinfo{journal}{arXiv:1106.0280}  (\bibinfo{year}{2011}),
  \eprint{1106.0280}.

\bibitem[{\citenamefont{M.}(2014)}]{lanzagorta2014}
\bibinfo{author}{\bibfnamefont{L.}~\bibnamefont{M.}},
  \emph{\bibinfo{title}{{Quantum Information in Gravitational Fields}}}
  (\bibinfo{publisher}{Morgan {\&} Claypool Publishers}, \bibinfo{year}{2014}).

\bibitem[{\citenamefont{Hawking}(1976)}]{hawking1976}
\bibinfo{author}{\bibfnamefont{S.~W.} \bibnamefont{Hawking}},
  \bibinfo{journal}{Physical Review D} \textbf{\bibinfo{volume}{14}},
  \bibinfo{pages}{2460} (\bibinfo{year}{1976}).

\bibitem[{\citenamefont{Almheiri et~al.}(2013)\citenamefont{Almheiri, Marolf,
  Polchinski, and Sully}}]{almheiri2013}
\bibinfo{author}{\bibfnamefont{A.}~\bibnamefont{Almheiri}},
  \bibinfo{author}{\bibfnamefont{D.}~\bibnamefont{Marolf}},
  \bibinfo{author}{\bibfnamefont{J.}~\bibnamefont{Polchinski}},
  \bibnamefont{and} \bibinfo{author}{\bibfnamefont{J.}~\bibnamefont{Sully}},
  \bibinfo{journal}{Journal of High Energy Physics}
  \textbf{\bibinfo{volume}{2013}}, \bibinfo{pages}{062} (\bibinfo{year}{2013}).

\bibitem[{\citenamefont{Page}(2005)}]{page2005}
\bibinfo{author}{\bibfnamefont{D.~N.} \bibnamefont{Page}},
  \bibinfo{journal}{New Journal of Physics} \textbf{\bibinfo{volume}{7}},
  \bibinfo{pages}{203} (\bibinfo{year}{2005}).

\bibitem[{\citenamefont{Giddings}(2016)}]{giddings2016}
\bibinfo{author}{\bibfnamefont{S.~B.} \bibnamefont{Giddings}},
  \bibinfo{journal}{Physics Letters B} \textbf{\bibinfo{volume}{754}},
  \bibinfo{pages}{39} (\bibinfo{year}{2016}).

\bibitem[{\citenamefont{Hod}(2016)}]{hod2016}
\bibinfo{author}{\bibfnamefont{S.}~\bibnamefont{Hod}},
  \bibinfo{journal}{Physics Letters B} \textbf{\bibinfo{volume}{757}},
  \bibinfo{pages}{121} (\bibinfo{year}{2016}).

\bibitem[{\citenamefont{Dey et~al.}(2017)\citenamefont{Dey, Liberati, and
  Pranzetti}}]{dey2017}
\bibinfo{author}{\bibfnamefont{R.}~\bibnamefont{Dey}},
  \bibinfo{author}{\bibfnamefont{S.}~\bibnamefont{Liberati}}, \bibnamefont{and}
  \bibinfo{author}{\bibfnamefont{D.}~\bibnamefont{Pranzetti}},
  \bibinfo{journal}{Physics Letters B} \textbf{\bibinfo{volume}{774}},
  \bibinfo{pages}{308} (\bibinfo{year}{2017}).

\bibitem[{\citenamefont{Dey et~al.}(2019)\citenamefont{Dey, Liberati,
  Mirzaiyan, and D.~Pranzetti}}]{dey2019}
\bibinfo{author}{\bibfnamefont{R.}~\bibnamefont{Dey}},
  \bibinfo{author}{\bibfnamefont{S.}~\bibnamefont{Liberati}},
  \bibinfo{author}{\bibfnamefont{Z.}~\bibnamefont{Mirzaiyan}},
  \bibnamefont{and}
  \bibinfo{author}{\bibfnamefont{D.}~\bibnamefont{D.~Pranzetti}},
  \bibinfo{journal}{Physics Letters B} \textbf{\bibinfo{volume}{797}},
  \bibinfo{pages}{134828} (\bibinfo{year}{2019}).

\bibitem[{\citenamefont{Zhang and Zhang}(2025)}]{zhang2025quantum}
\bibinfo{author}{\bibfnamefont{H.}~\bibnamefont{Zhang}} \bibnamefont{and}
  \bibinfo{author}{\bibfnamefont{B.}~\bibnamefont{Zhang}},
  \bibinfo{journal}{Physical Review D} \textbf{\bibinfo{volume}{111}},
  \bibinfo{pages}{085007} (\bibinfo{year}{2025}).

\bibitem[{\citenamefont{Peres et~al.}(2002)\citenamefont{Peres, Scudo, and
  R.}}]{peres2002}
\bibinfo{author}{\bibfnamefont{A.}~\bibnamefont{Peres}},
  \bibinfo{author}{\bibfnamefont{P.~F.} \bibnamefont{Scudo}}, \bibnamefont{and}
  \bibinfo{author}{\bibfnamefont{T.~D.} \bibnamefont{R.}},
  \bibinfo{journal}{Physical Review Letters} \textbf{\bibinfo{volume}{88}},
  \bibinfo{pages}{230402} (\bibinfo{year}{2002}).

\bibitem[{\citenamefont{Peres}(1996)}]{peres1996}
\bibinfo{author}{\bibfnamefont{A.}~\bibnamefont{Peres}},
  \bibinfo{journal}{Physical Review Letters} \textbf{\bibinfo{volume}{77}},
  \bibinfo{pages}{1413} (\bibinfo{year}{1996}).

\bibitem[{\citenamefont{Ollivier and Zurek}(2001)}]{ollivier2001}
\bibinfo{author}{\bibfnamefont{H.}~\bibnamefont{Ollivier}} \bibnamefont{and}
  \bibinfo{author}{\bibfnamefont{W.~H.} \bibnamefont{Zurek}},
  \bibinfo{journal}{Physical Review Letters} \textbf{\bibinfo{volume}{88}},
  \bibinfo{pages}{017901} (\bibinfo{year}{2001}).

\bibitem[{\citenamefont{Luo and Fu}(2011)}]{luo2011}
\bibinfo{author}{\bibfnamefont{S.}~\bibnamefont{Luo}} \bibnamefont{and}
  \bibinfo{author}{\bibfnamefont{S.}~\bibnamefont{Fu}},
  \bibinfo{journal}{Physical Review Letters} \textbf{\bibinfo{volume}{106}},
  \bibinfo{pages}{120401} (\bibinfo{year}{2011}).

\bibitem[{\citenamefont{Kaczmarek and Szcz{\c
  e}{\'s}niak}(2024)}]{kaczmarek2024}
\bibinfo{author}{\bibfnamefont{A.~Z.} \bibnamefont{Kaczmarek}}
  \bibnamefont{and} \bibinfo{author}{\bibfnamefont{D.}~\bibnamefont{Szcz{\c
  e}{\'s}niak}}, \bibinfo{journal}{Physics Letters B}
  \textbf{\bibinfo{volume}{848}}, \bibinfo{pages}{138364}
  (\bibinfo{year}{2024}).

\bibitem[{\citenamefont{Kaczmarek et~al.}(2025)\citenamefont{Kaczmarek, Szcz{\c
  e}{\'s}niak, B{\c a}k, and Szcz{\c e}{\'s}niak}}]{kaczmarek2025}
\bibinfo{author}{\bibfnamefont{A.~Z.} \bibnamefont{Kaczmarek}},
  \bibinfo{author}{\bibfnamefont{D.}~\bibnamefont{Szcz{\c e}{\'s}niak}},
  \bibinfo{author}{\bibfnamefont{Z.}~\bibnamefont{B{\c a}k}}, \bibnamefont{and}
  \bibinfo{author}{\bibfnamefont{R.}~\bibnamefont{Szcz{\c e}{\'s}niak}},
  \bibinfo{journal}{Physics Letters B} \textbf{\bibinfo{volume}{868}},
  \bibinfo{pages}{139683} (\bibinfo{year}{2025}).

\bibitem[{\citenamefont{Liu et~al.}(2024)\citenamefont{Liu, Long, and
  He}}]{liu2024quantum}
\bibinfo{author}{\bibfnamefont{C.}~\bibnamefont{Liu}},
  \bibinfo{author}{\bibfnamefont{Z.}~\bibnamefont{Long}}, \bibnamefont{and}
  \bibinfo{author}{\bibfnamefont{Q.}~\bibnamefont{He}},
  \bibinfo{journal}{Physics Letters B} \textbf{\bibinfo{volume}{857}},
  \bibinfo{pages}{138991} (\bibinfo{year}{2024}).

\bibitem[{\citenamefont{Zhang et~al.}(2025)\citenamefont{Zhang, Li, Song, Ye,
  and Wang}}]{zhang2025entanglement}
\bibinfo{author}{\bibfnamefont{S.}~\bibnamefont{Zhang}},
  \bibinfo{author}{\bibfnamefont{L.}~\bibnamefont{Li}},
  \bibinfo{author}{\bibfnamefont{X.}~\bibnamefont{Song}},
  \bibinfo{author}{\bibfnamefont{L.}~\bibnamefont{Ye}}, \bibnamefont{and}
  \bibinfo{author}{\bibfnamefont{D.}~\bibnamefont{Wang}},
  \bibinfo{journal}{Physics Letters B} \textbf{\bibinfo{volume}{868}},
  \bibinfo{pages}{139648} (\bibinfo{year}{2025}).

\bibitem[{\citenamefont{Zhang and Wang}(2026)}]{zhang2026quantum}
\bibinfo{author}{\bibfnamefont{S.}~\bibnamefont{Zhang}} \bibnamefont{and}
  \bibinfo{author}{\bibfnamefont{D.}~\bibnamefont{Wang}},
  \bibinfo{journal}{Physics Letters B} \textbf{\bibinfo{volume}{875}},
  \bibinfo{pages}{140354} (\bibinfo{year}{2026}).

\bibitem[{\citenamefont{Liu et~al.}(2026)\citenamefont{Liu, Wen, and
  Wang}}]{liu2026quantum}
\bibinfo{author}{\bibfnamefont{X.}~\bibnamefont{Liu}},
  \bibinfo{author}{\bibfnamefont{C.}~\bibnamefont{Wen}}, \bibnamefont{and}
  \bibinfo{author}{\bibfnamefont{J.}~\bibnamefont{Wang}},
  \bibinfo{journal}{Physics Letters B} \textbf{\bibinfo{volume}{873}},
  \bibinfo{pages}{140185} (\bibinfo{year}{2026}).

\bibitem[{\citenamefont{Ming et~al.}(2026)\citenamefont{Ming, Lu, Xu, Dong,
  Fang, Hu, Yu, Yang, and Wang}}]{ming2026nonlocal}
\bibinfo{author}{\bibfnamefont{F.}~\bibnamefont{Ming}},
  \bibinfo{author}{\bibfnamefont{T.}~\bibnamefont{Lu}},
  \bibinfo{author}{\bibfnamefont{Z.}~\bibnamefont{Xu}},
  \bibinfo{author}{\bibfnamefont{L.}~\bibnamefont{Dong}},
  \bibinfo{author}{\bibfnamefont{B.}~\bibnamefont{Fang}},
  \bibinfo{author}{\bibfnamefont{X.}~\bibnamefont{Hu}},
  \bibinfo{author}{\bibfnamefont{Y.}~\bibnamefont{Yu}},
  \bibinfo{author}{\bibfnamefont{H.}~\bibnamefont{Yang}}, \bibnamefont{and}
  \bibinfo{author}{\bibfnamefont{D.}~\bibnamefont{Wang}},
  \bibinfo{journal}{Physics Letters B} \textbf{\bibinfo{volume}{874}},
  \bibinfo{pages}{140242} (\bibinfo{year}{2026}).

\bibitem[{\citenamefont{Ming-Liang and Heng}(2012)}]{ming-liang2012}
\bibinfo{author}{\bibfnamefont{H.}~\bibnamefont{Ming-Liang}} \bibnamefont{and}
  \bibinfo{author}{\bibfnamefont{F.}~\bibnamefont{Heng}},
  \bibinfo{journal}{Annals of Physics} \textbf{\bibinfo{volume}{327}},
  \bibinfo{pages}{2343} (\bibinfo{year}{2012}).

\bibitem[{\citenamefont{Daki{\'c} et~al.}(2010)\citenamefont{Daki{\'c}, Vedral,
  and Brukner}}]{dakic2010}
\bibinfo{author}{\bibfnamefont{B.}~\bibnamefont{Daki{\'c}}},
  \bibinfo{author}{\bibfnamefont{V.}~\bibnamefont{Vedral}}, \bibnamefont{and}
  \bibinfo{author}{\bibfnamefont{C.}~\bibnamefont{Brukner}},
  \bibinfo{journal}{Physical Review Letters} \textbf{\bibinfo{volume}{105}},
  \bibinfo{pages}{190502} (\bibinfo{year}{2010}).

\bibitem[{\citenamefont{Kaczmarek et~al.}(2023)\citenamefont{Kaczmarek, Szcz{\c
  e}{\'s}niak, and Kais}}]{kaczmarek2023}
\bibinfo{author}{\bibfnamefont{A.}~\bibnamefont{Kaczmarek}},
  \bibinfo{author}{\bibfnamefont{D.}~\bibnamefont{Szcz{\c e}{\'s}niak}},
  \bibnamefont{and} \bibinfo{author}{\bibfnamefont{S.}~\bibnamefont{Kais}},
  \bibinfo{journal}{Universe} \textbf{\bibinfo{volume}{9}},
  \bibinfo{pages}{199} (\bibinfo{year}{2023}).

\bibitem[{\citenamefont{Hartle and Hawking}(1976)}]{hartlehawking1976}
\bibinfo{author}{\bibfnamefont{J.~B.} \bibnamefont{Hartle}} \bibnamefont{and}
  \bibinfo{author}{\bibfnamefont{S.~W.} \bibnamefont{Hawking}},
  \bibinfo{journal}{Physical Review D} \textbf{\bibinfo{volume}{13}},
  \bibinfo{pages}{2188} (\bibinfo{year}{1976}).

\bibitem[{\citenamefont{Tolman and Ehrenfest}(1930)}]{tolman1930}
\bibinfo{author}{\bibfnamefont{R.~C.} \bibnamefont{Tolman}} \bibnamefont{and}
  \bibinfo{author}{\bibfnamefont{P.}~\bibnamefont{Ehrenfest}},
  \bibinfo{journal}{Physical Review} \textbf{\bibinfo{volume}{36}},
  \bibinfo{pages}{1791} (\bibinfo{year}{1930}).

\bibitem[{\citenamefont{Eune and Kim}(2019)}]{eune2019}
\bibinfo{author}{\bibfnamefont{M.}~\bibnamefont{Eune}} \bibnamefont{and}
  \bibinfo{author}{\bibfnamefont{W.}~\bibnamefont{Kim}},
  \bibinfo{journal}{Physics Letters B} \textbf{\bibinfo{volume}{798}},
  \bibinfo{pages}{135020} (\bibinfo{year}{2019}).

\bibitem[{\citenamefont{Navarro-Salas and Fabbri}(2005)}]{navarro2005}
\bibinfo{author}{\bibfnamefont{J.}~\bibnamefont{Navarro-Salas}}
  \bibnamefont{and} \bibinfo{author}{\bibfnamefont{A.}~\bibnamefont{Fabbri}},
  \emph{\bibinfo{title}{{Modeling Black Hole Evaporation}}}
  (\bibinfo{publisher}{World Scientific}, \bibinfo{year}{2005}).

\bibitem[{\citenamefont{Pan and Jing}(2008)}]{pan2008}
\bibinfo{author}{\bibfnamefont{Q.}~\bibnamefont{Pan}} \bibnamefont{and}
  \bibinfo{author}{\bibfnamefont{J.}~\bibnamefont{Jing}},
  \bibinfo{journal}{Physical Review D} \textbf{\bibinfo{volume}{78}},
  \bibinfo{pages}{065015} (\bibinfo{year}{2008}).

\bibitem[{\citenamefont{IFuentes-Schuller and Mann}(2005)}]{fuentes2005}
\bibinfo{author}{\bibfnamefont{I.}~\bibnamefont{IFuentes-Schuller}}
  \bibnamefont{and} \bibinfo{author}{\bibfnamefont{R.~B.} \bibnamefont{Mann}},
  \bibinfo{journal}{Physical Review Letters} \textbf{\bibinfo{volume}{95}},
  \bibinfo{pages}{120404} (\bibinfo{year}{2005}).

\bibitem[{\citenamefont{He et~al.}(2016)\citenamefont{He, Xu, and Ye}}]{he2016}
\bibinfo{author}{\bibfnamefont{J.}~\bibnamefont{He}},
  \bibinfo{author}{\bibfnamefont{S.}~\bibnamefont{Xu}}, \bibnamefont{and}
  \bibinfo{author}{\bibfnamefont{L.}~\bibnamefont{Ye}},
  \bibinfo{journal}{Physics Letters B} \textbf{\bibinfo{volume}{756}},
  \bibinfo{pages}{278} (\bibinfo{year}{2016}).

\bibitem[{\citenamefont{Ge and Kim}(2008)}]{ge2008}
\bibinfo{author}{\bibfnamefont{X.~H.} \bibnamefont{Ge}} \bibnamefont{and}
  \bibinfo{author}{\bibfnamefont{S.~P.} \bibnamefont{Kim}},
  \bibinfo{journal}{Class. Quantum Gravity} \textbf{\bibinfo{volume}{25}},
  \bibinfo{pages}{075011} (\bibinfo{year}{2008}).

\bibitem[{\citenamefont{Alsing and Milburn}(2003)}]{alsing2003}
\bibinfo{author}{\bibfnamefont{P.~M.} \bibnamefont{Alsing}} \bibnamefont{and}
  \bibinfo{author}{\bibfnamefont{G.~J.} \bibnamefont{Milburn}},
  \bibinfo{journal}{Physical Review Letters} \textbf{\bibinfo{volume}{91}},
  \bibinfo{pages}{180404} (\bibinfo{year}{2003}).

\bibitem[{\citenamefont{Bruschi et~al.}(2010)\citenamefont{Bruschi, Louko,
  Mart\'{\i}n-Mart\'{\i}nez, Dragan, and Fuentes}}]{bruschi2010}
\bibinfo{author}{\bibfnamefont{D.~E.} \bibnamefont{Bruschi}},
  \bibinfo{author}{\bibfnamefont{J.}~\bibnamefont{Louko}},
  \bibinfo{author}{\bibfnamefont{E.}~\bibnamefont{Mart\'{\i}n-Mart\'{\i}nez}},
  \bibinfo{author}{\bibfnamefont{A.}~\bibnamefont{Dragan}}, \bibnamefont{and}
  \bibinfo{author}{\bibfnamefont{I.}~\bibnamefont{Fuentes}},
  \bibinfo{journal}{Physical Review A} \textbf{\bibinfo{volume}{82}},
  \bibinfo{pages}{042332} (\bibinfo{year}{2010}).

\bibitem[{\citenamefont{Hu et~al.}(2018)\citenamefont{Hu, Hu, Wang, Peng,
  Zhang, and Fan}}]{HU2018}
\bibinfo{author}{\bibfnamefont{M.}~\bibnamefont{Hu}},
  \bibinfo{author}{\bibfnamefont{X.}~\bibnamefont{Hu}},
  \bibinfo{author}{\bibfnamefont{J.}~\bibnamefont{Wang}},
  \bibinfo{author}{\bibfnamefont{Y.}~\bibnamefont{Peng}},
  \bibinfo{author}{\bibfnamefont{Y.~R.} \bibnamefont{Zhang}}, \bibnamefont{and}
  \bibinfo{author}{\bibfnamefont{H.}~\bibnamefont{Fan}},
  \bibinfo{journal}{Physics Reports} \textbf{\bibinfo{volume}{762-764}},
  \bibinfo{pages}{1} (\bibinfo{year}{2018}).

\bibitem[{\citenamefont{Zettili}(2009)}]{zettili2009quantum}
\bibinfo{author}{\bibfnamefont{N.}~\bibnamefont{Zettili}},
  \emph{\bibinfo{title}{Quantum Mechanics: Concepts and Applications}}
  (\bibinfo{publisher}{John Wiley \& Sons}, \bibinfo{year}{2009}).

\bibitem[{\citenamefont{Richter and Omar}(2015)}]{richter2015degradation}
\bibinfo{author}{\bibfnamefont{B.}~\bibnamefont{Richter}} \bibnamefont{and}
  \bibinfo{author}{\bibfnamefont{Y.}~\bibnamefont{Omar}},
  \bibinfo{journal}{Physical Review A} \textbf{\bibinfo{volume}{92}},
  \bibinfo{pages}{022334} (\bibinfo{year}{2015}).

\bibitem[{\citenamefont{Brown et~al.}(2012)\citenamefont{Brown, Cormier,
  Mart\'{\i}n-Mart\'{\i}nez, and Mann}}]{brown2012}
\bibinfo{author}{\bibfnamefont{E.~G.} \bibnamefont{Brown}},
  \bibinfo{author}{\bibfnamefont{K.}~\bibnamefont{Cormier}},
  \bibinfo{author}{\bibfnamefont{E.}~\bibnamefont{Mart\'{\i}n-Mart\'{\i}nez}},
  \bibnamefont{and} \bibinfo{author}{\bibfnamefont{R.~B.} \bibnamefont{Mann}},
  \bibinfo{journal}{Physical Review A} \textbf{\bibinfo{volume}{86}},
  \bibinfo{pages}{032108} (\bibinfo{year}{2012}).

\bibitem[{\citenamefont{Tian and Jing}(2013)}]{tian2013}
\bibinfo{author}{\bibfnamefont{Z.}~\bibnamefont{Tian}} \bibnamefont{and}
  \bibinfo{author}{\bibfnamefont{J.}~\bibnamefont{Jing}},
  \bibinfo{journal}{Annals of Physics} \textbf{\bibinfo{volume}{333}},
  \bibinfo{pages}{76} (\bibinfo{year}{2013}).

\end{thebibliography}

\end{document}